# Is Coronavirus-Related Research Becoming More Interdisciplinary? A Perspective of Co-occurrence Analysis and Diversity Measure of Scientific Articles

Yi Zhao, Lifan Liu, Chengzhi Zhang·

Department of Information Management, Nanjing University of Science and Technology, Nanjing, China, 210094
yizhao93@njust.edu.cn, liulf@njust.edu.cn, zhangcz@njust.edu.cn

**Abstract.** The outbreak of coronavirus disease 2019 (COVID-19) has had a significant repercussion on the health, economy, politics and environment, making coronavirus-related issues more complicated and difficult to solve adequately by relying on a single field. Interdisciplinary research can provide an effective solution to complex issues in the related field of coronavirus. However, whether coronavirus related research becomes more interdisciplinary still needs corroboration. In this study, we investigate interdisciplinary status of the coronavirus-related fields via the COVID-19 Open Research Dataset (CORD-19). To this end, we calculate bibliometric indicators of interdisciplinarity and a co-occurrence analysis method. The results show that co-occurrence relationships between cited disciplines have evolved dynamically over time. The two types of co-occurrence relationships, Immunology and Microbiology & Medicine and Chemical Engineering & Chemistry, last for a long time in this field during 1990-2020. Moreover, the number of disciplines cited by coronavirus-related research increases, whereas the distribution of disciplines is uneven, and this field tends to focus on several dominant disciplines such as Medicine, Immunology and Microbiology, Biochemistry, Genetics and Molecular Biology. We also gauge the disciplinary diversity of COVID-19 related papers published from January to December 2020; the disciplinary variety shows an upward trend, while the degree of disciplinary balance shows a downward trend. Meanwhile, the comprehensive index $^2D^s$ demonstrates that the degree of interdisciplinarity in coronavirus field decreases between 1990 and 2019, but it increases in 2020. The results help to map the interdisciplinarity of coronavirus-related research, gaining insight into the degree and history of interdisciplinary cooperation.

# 1 Introduction

In December 2019, an infectious disease called Coronavirus Disease 2019 (COVID-19) caused by an emerging coronavirus, SARS-CoV-2, spread globally. As the number of confirmed cases rapidly increased, the World Health Organization (WHO) declared the COVID-19 outbreak as a global pandemic, on March 11, 2020[1]. As of 7 September 2020, there are 26,981,755 confirmed cases and 882,053 deaths worldwide[2]. The COVID-19 pandemic has had a massive impact on global health, the economy and politics, which is becoming an unprecedented challenge facing humanity in the 21st century.

In the context of the COVID-19 pandemic, numerous scholars have changed the focus of their research work, using their professional knowledge to conduct research in multiple subjects such as disease prevention, diagnosis, treatment, and vaccine development (Homolak et al. 2020). Real-world problems in science and technology cannot be solved adequately using the knowledge from any single discipline (Bruine de Bruin and Morgan 2019), and COVID-19 related issues are no exception. Interdisciplinary research has thus become an effective way to tackle these complex global problems.

Interdisciplinary research (IDR) was defined as a mode of research that integrated knowledge from different discipline, and it has been a common practice in scientific investigation (Qin et al. 1997; Alan L Porter et al. 2006). Quantitative measurement of interdisciplinarity helps to understand the evolution of disciplines in a specific field and facilitates the knowledge integration of different disciplines. In addition, exploring the pattern of disciplinary co-occurrence can provide new insights into the choice of the forthcoming advanced directions. Scientists have called for an interdisciplinary collaboration that is required for solving COIVD-19 related issues (Bontempi et al. 2020). In this context, a question arises: is coronavirus related research becoming more interdisciplinary after the outbreak of the COVID-19 pandemic? Moreover, few studies have investigated the interdisciplinary characteristics of coronavirus-related research. For the reasons noted above, this study aims to examine the degree of interdisciplinarity of coronavirus-related research, and explore the evolution of disciplinary co-occurrence relationships.

---

[1] Available at: https://www.who.int/dg/speeches/detail/who-director-general-s-opening-remarks-at-the-media-briefing-on-covid-19---11-march-2020. Retrieved on 7 September 2020.

[2] Available at: https://www.arcgis.com/apps/opsdashboard/index.html#/bda7594740fd40299423467b48e9ecf6. Retrieved on 7 September 2020.

Specifically, the main research questions of this article are as follows:

*RQ1*: What disciplines are involved in coronavirus-related research? which disciplines are being cited by coronavirus related research?

*RQ2*: How does the disciplinary co-occurrence relationships of the references evolve over time?

*RQ3*: What's the degree of interdisciplinarity in coronavirus-related research fields, and is coronavirus-related research becoming more interdisciplinary?

## 2    Related Work

### 2.1    Interdisciplinary research

Scholars have extensively investigated the interdisciplinary phenomenon in scientific publications using various methods, and a great deal of research has been published. These previous studies have largely focused on the measurement of interdisciplinary degree and the evolution of interdisciplinary collaboration (Abramo et al. 2018; Hu and Zhang 2018; Su and Moaniba 2017).

With respect to the measurement of interdisciplinary degree, most of these prior studies have mainly used the index to measure the degree of interdisciplinarity, including the Stirling's diversity index (Stirling 2007), Shannon entropy and the Gini coefficient (Leydesdorff and Rafols 2011), Disciplinary Diversity Index (Alan L. Porter and Rafols 2009), DIV index (Leydesdorff et al. 2019), etc. Additionally, L. Zhang et al. (2015) used a Hill-type true diversity indicator, namely $^2D^s$, to measure interdisciplinary research, and found that $^2D^s$ perform better than the Rao-Stirling indicator. Furthermore, scientists have made great attempts to analyze the evolution of interdisciplinary collaboration and provided several practical ways, such as co-authorship network analysis (C. Zhang et al. 2018), co-citation network analysis (Chi and Young 2013) and co-occurrence analysis (Hu and Zhang 2017; J. Xu et al. 2018). More importantly, Leydesdorff and Goldstone (2014) utilized newly available software called Cortext to map the relationship between disciplines, and found that it can work well for dynamic visualization and animation of complex contexts. Additionally, the study of interdisciplinary nature has covered a wide spectrum of fields, such as information behavior (Deng and Xia 2020; Jamali and Nicholas 2010), biotechnology (Alan L. Porter and Rafols 2009), library and information science (Chang 2018), big data (Hu and Zhang 2017), precision medicine (X. Xu et al. 2021). However, as argued by J. Xu et al. (2018), scientific problems in health and medicine are also becoming more complicated, and interdisciplinary research is needed to facilitate innovation to tackle them. Nevertheless, the interdisciplinarity of medicine still remains to be explored.

Citation analysis and co-authorship analysis are two representative bibliometric methods, which are extensively used for interdisciplinary degree measurement in multiple research domains. Citation analysis utilizes the discipline classification system to classify citations into different pre-defined categories, and then explores the interactions between disciplines in different research domains (Rafols and Meyer 2010), but it cannot reveal the interdisciplinary degree of one research field. Co-authorship analysis adopts the cooperation between scholars specialized in different research areas to measure the interdisciplinarity of a particular field (Huang and Chang 2011), but it is an indirect way to capture the knowledge integration phenomenon. Diversity measures serve as an alternative method to complement existing bibliometric analysis methods. $^2D^S$ index has achieved good performance in discrimination and accuracy (Abramo et al. 2018; Deng and Xia 2020). Moreover, variety, balance and disparity are three aspects of interdisciplinarity (Stirling 2007). To provide a comprehensive panorama of interdisciplinarity in coronavirus field, all indexes are used in this study.

### 2.2 Bibliometric analysis of coronavirus-related research

The COVID-19 pandemic posed a tremendous challenge to the global research community, and researchers rushed to find solutions to the catastrophic crisis, which also contributed to an increase in the number of coronavirus-related publications at a spectacular rate (Y. Zhang et al. 2021). Coronavirus-related publications provide a rich resource for further research, and there are many efforts to utilize these publication data to conduct bibliometric analysis. Dehghanbanadaki et al. (2020) analyzed 923 COVID-19-related articles downloaded from the Scopus database, and the results showed that China is the most prolific country in the sample, with China and the United States exhibiting the strongest collaboration relationship. Belli et al. (2020) collected 18,875 articles indexed in the Web of Science, identified the most productive authors, organizations, countries in the coronavirus related publications. L. Zhang et al. (2020) discussed the response of academic community to six international public health emergencies since 2000, including SARS, influenza A (H1N1), ebola, zika and COVID-19, and they also presented a preliminary analysis of academic responses to COVID-19. Mao et al. (2020) explored the global status and trends of coronavirus research, and found that the USA is the most prolific country in coronavirus publications. Several scholars have also conducted studies on the scientific collaboration (Cai et al. 2021). Homolak et al. (2020) evaluated the quality of scientific collaboration during the COVID-19 pandemic, and the preliminary analysis results revealed significant problems with apparent insufficient collaboration. Fry et al. (2020) investigated whether the COVID-19 pandemic accelerated or reversed international collaboration, and reviewed the articles published in the first months of the COVID-19 pandemic. The results indicated that COVID-19 research had a smaller team size and involved fewer participants from different countries, when compared to pre-COVID-19 coronavirus research. Additional, a few studies have explored the relationship between policy and science (Yin et al. 2021). Wu et al. (2021) used bibliometric methods to characterize the patterns of China's policy against COVID-19.

As task difficulty in the coronavirus field increases, it is critical to integrate knowledge from different disciplines, and many researchers have also called on the scientific community to conduct interdisciplinary research to address the problems.

Previous studies generally focused on one or more aspects of bibliometric issues such as the growth of COVID-19 publications, international collaboration, national contributions, etc. However, whether coronavirus-related research becomes more interdisciplinary remains unclear. At the same time, quantitative studies on the level of knowledge integration and disciplinary co-occurrence patterns of coronavirus-related research are still lacking.

## 3   Methodology

The framework of this study is illustrated in Figure 1, which mainly consists of three parts: data preparation, data processing and data analysis. First, data preparation involves two steps, data collecting and data preprocessing. Then, we need to tidy up the disciplines of each qualified article, and calculate the degree of interdisciplinarity in the data processing phase. Finally, we analyze the disciplinary distribution of references in coronavirus-related research, the disciplinary co-occurrence patterns and its evolution, and the degree of interdisciplinarity.

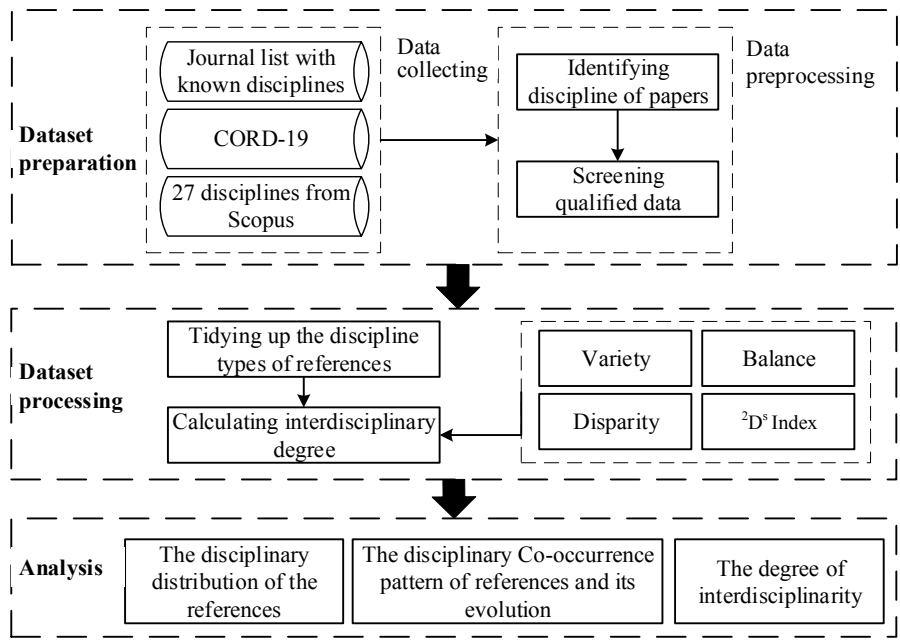

Figure 1 Framework of this study

### 3.1   Data

Dataset preparation, as shown in Figure 2, has two main steps. First, we need to collect data from CORD-19. Second, to analyze the interdisciplinarity of coronavirus-related research, we must identify the disciplines of papers and screen the qualified data for further analysis.

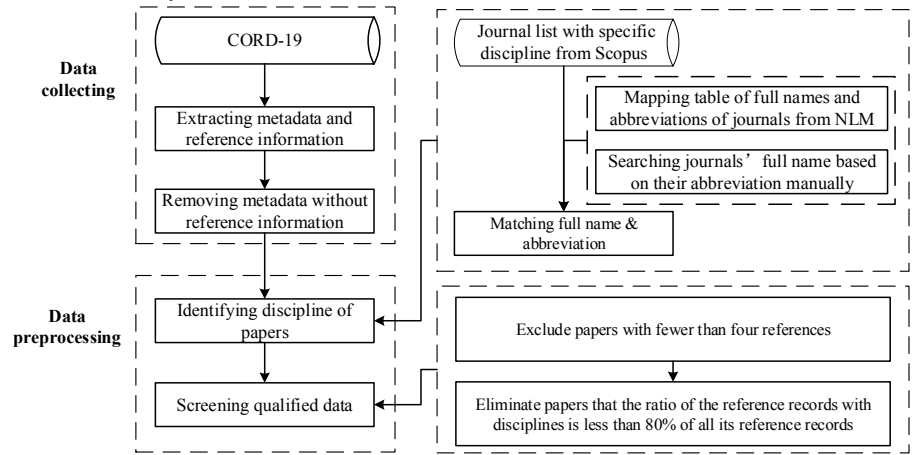

Figure 2 Framework of data preparation

### 3.1.1 Data collecting

The data used in this paper, named CORD-19, were released by the Allen Institute for AI in partnership with other companies[3], which is a large and updated collection of publications on COVID-19 and related coronaviruses. Papers in CORD-19 are collected from several database, including PubMed, Elsevier, Springer Nature, the WHO's Covid-19 Database and preprint servers. The following query was used in searches in the Title/Abstract of each paper in the respective databases: "COVID" OR "COVID-19" OR "Coronavirus" OR "Corona virus" OR "2019-nCoV" OR "SARS-CoV" OR "MERS-CoV" OR "Severe Acute Respiratory Syndrome" OR "Middle East Respiratory Syndrome"(Wang et al. 2020).

We parsed all publication metadata (title, abstract, publication time, author, journal, reference, etc.) from CORD-19 (JSON format) and stored it in a local MySQL database. Here we collected 440,306 publications published from January 1, 1990 to December 31, 2020, and restricted the publication types to "Article" and "Review" (The same criteria applied to the references). Of these, 288,719 articles without corresponding references are removed from the dataset. Therefore, 151,587 articles with 5,517,523 corresponding references comprise our dataset for further data preprocessing.

### 3.1.2 Data preprocessing

Data preprocessing has two core sub-steps. First, we need to classify references into pre-defined discipline categories. Then, to produce a credible and valid result, we use some criteria to screen qualified paper records.

（1）**Identifying discipline of references**

---

[3] COVID-19 Open Research Dataset (CORD-19). 2021. Version 2021-07-12. Available at: https://pages.semanticscholar.org/coronavirus-research. Accessed 2021-07-12.

The prerequisite to this study is the identification of the article's discipline, and mainstream identification methods can be grouped into two categories: top-down and bottom-up approaches (Liu et al. 2012). Top-down approaches classify each article into the pre-defined categories, and this approach is adopted by many bibliographic databases, such as Web of Science (Leydesdorff et al. 2019), Scopus (C. Zhang et al. 2021), Microsoft Academic Graph (Zhai; et al. 2021), etc. Bottom-up approaches aim to explore the intrinsic structure of scholarly networks and extract the clusters as disciplines. Some bottom-up approaches are typically based on the complex network, such as co-word networks (Bu et al. 2021) and co-citation networks (Zeng et al. 2019). Both approaches have their pros and cons; top-down approaches are popular and easy to use, but they cannot capture the fine-grained research topics. Bottom-up approaches enable us to investigate the dynamics of discipline, but the disadvantage is also obvious: the gold standard is lacking and the accuracy of the clustering result is hard to verify. Given that top-down approaches are more mature and widely used in scientometrics (Rafols and Meyer 2010; Leydesdorff et al. 2012), we chose top-down approaches to identify the article's discipline in our study. Specifically, we assign the discipline to each article according to the discipline classification system of Scopus[4]. More precisely, we identify the disciplines of the article based on the disciplines of the journal (C. Zhang et al. 2021). Scopus implements a more fine-grained discipline category, which contains 27 first-level disciplines. More importantly, Scopus provides a journal list that includes over 23,000 journals with one or more disciplines.

In this study, we got a total of 520,993 journal titles, 11,967 of which were from papers (citing) and 509,026 of which were from references. The full names of the journal titles and various abbreviations are included. For example, the journal of *Proceedings of the National Academy of Sciences of the United States of America* has multiple abbreviation forms, such as *PNAS*, *proc natl acad sci*, *proc natl acad sci usa*, *proc. natl. acad. sci. usa*, etc.

First, we downloaded a mapping table between journal full names and journal title abbreviations from the National Library of Medicine[5]. Based on the mapping table, we transformed the abbreviations of some journal names to their full names. Then, we matched all the records (151,587 articles and 5,517,523 references) to the journal list from Scopus. A proportion of records (69% of articles and 61% of references) can be matched to their disciplines. For the remaining journals with unknown discipline from papers(citing) and from references, when the frequency of journal titles is greater than 1 and 29, respectively, we manually retrieved the full names of journal titles based on their abbreviations. Further, the retrieved results were cross-checked in the group and submitted to an expert for final confirmation. Next, we matched remaining records to the journal list from Scopus. Therefore, 98.79% of papers (149756/151587) and 76.9% of references (4,243,484/5,517,523) were assigned one or more disciplines. It is noteworthy that only the journals indexed by Scopus could be assigned to one or more disciplines.

（2）**Screening qualified data**

In order to guarantee the reliability of results, we removed the papers(citing) with fewer than four references. We did this for two reasons. First, we use the diversity of references to measure the diversity of research areas. Second, this criterion helps filter out incomplete records. Additionally, based on our previous experience, we eliminated papers that the ratio of the reference records with disciplines is less than 80% of all its reference records (C. Zhang et al. 2021). Finally, 48,469 articles and 2,224,048 references comprise our final data set (Detailed information is shown in Table 1). In addition, we searched "COVID" OR "COVID-19" OR "2019-nCoV" in the discipline-identifiable qualified data and retrieved all COVID-19 related articles published in 2020, and we obtained 21,354 articles and 639,995 references.

---

[4] Available at: https://www.scopus.com/
[5] Available at: https://www.nlm.nih.gov/

Table 1 The number of articles(citing) and corresponding references

| Data | # of citing papers | # of references |
| --- | --- | --- |
| Raw data | 440,306 | 5,517,523 |
| Discipline-identifiable data | 149,756 | 4,243,484 |
| *Discipline-identifiable qualified data | 48,469 | 2,224,048 |
| COVID-19 data (2020) | 21,354 | 639,995 |

**Note**: * qualified data means the articles have at least five references, and the ratio of the reference records with disciplines is more than 80% of all its reference records.

Figure 3 shows the number of coronavirus-related articles published each year covering 1990 to 2020. It can be readily seen that, before 2003, the number of articles was less than 300 per year. Since 2003, the number of coronavirus-related articles has steadily increased yearly. It is clear that a spike in published articles occurs in 2020 in response to COVID-19, and the total number of published articles exceeds the sum of the previous 30 years.

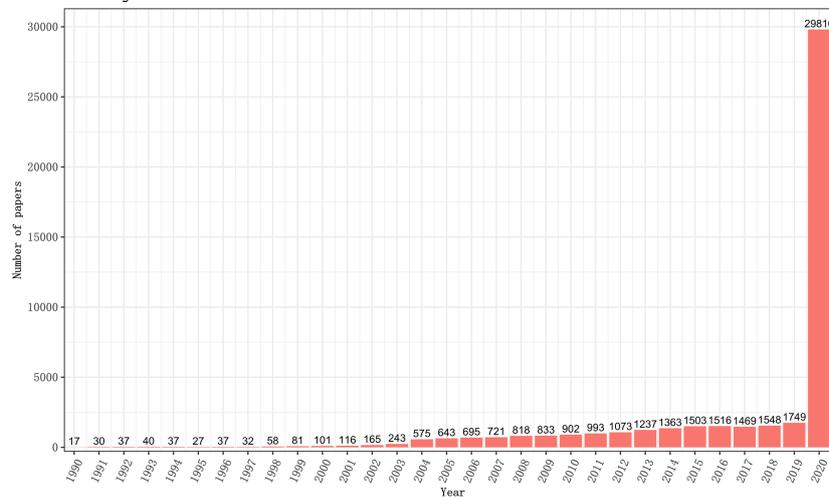

Figure 3 The number of coronavirus-related papers published each year covering 1990 to 2020

### 3.2 Analyzing the interdisciplinary status of coronavirus-related research

In this study, co-occurrence analysis was applied to explore the disciplinary co-occurrence patterns of references in coronavirus-related research. Additionally, diversity measures were applied to gauge the degree of interdisciplinarity of coronavirus-related research.

#### 3.2.1 Co-occurrence analysis of disciplines

We utilized a co-occurrence analysis approach (Small and Griffith 1974) to explore the co-occurrence relationships between disciplines. The co-occurrence analysis method can reveal the potential collaborative structure between two objects (keywords, authors, institutions, terminology, etc.), and has been widely used in multiple domains (Catala-Lopez et al. 2012; Grauwin and Jensen 2011; Su and Lee 2010). A large proportion of the literature on discipline co-occurrence analysis seems to have been based on the papers' SCs (Subject Categories) acquired from WoS (Web of Science). In contrast to previous research, this study investigated the interdisciplinarity of coronavirus-related research based on the disciplinary diversity of cited references (L. Zhang et al. 2015). Therefore, the measurement of co-occurrence patterns is transformed into the calculation of disciplinary co-occurrence of references. For example, article A cites two references, B and C, the reference B is assigned to two disciplines, *Medicine* and *Computer Science*, and the reference C is assigned to two disciplines, *Medicine* and *Decision Sciences*, and then *Medicine*, *Computer Science* and *Decision Sciences* represent the disciplinary co-occurrence pattern of article A.

To visualize the evolution of disciplinary co-occurrence patterns, we first collected the discipline of the references cited by coronavirus-related research, and then processed the data into a specific format, and finally, dynamic network analysis in the CorText[6] platform was employed for visualization. Specifically, Cortext allows for dynamic analysis of relationship components based on flow-diagrams of disciplinary co-occurrence relations, and it uses cosine algorithm to measure the similarity and Louvain-algorithm to decompose the community (Leydesdorff and Goldstone 2014; Blondel et al. 2008).

#### 3.2.2 Diversity measure of interdisciplinary degree

The key point of measuring the degree of interdisciplinarity is to characterize knowledge integration. and knowledge integration can be measured by investigating the breadth of a publication's references(Porter. et al. 2007). Figure 4 shows an example of the interdisciplinary phenomenon in COVID-19 related research from the citation perspective. It should be noted that the Figure 4 is just an operational example of the notion of interdisciplinarity. More specifically, it is a screenshot of a paper that studies the application of deep learning to COVID-19 drug repurposing (Pham et al. 2021). The paper was published in the journal of *Nature Machine Intelligence*, and it is assigned to Computer Science according to the journal classification system of the Scopus, while the cited references are assigned to Multidisciplinary, Medicine, Biochemistry, Genetics and Molecular Biology and Computer Science, so we can gauge diversity of papers based on the diversity of their references. In our study, we adopted the full-counting method for the classification system. As shown in Figure 4, the reference 5 was published in the journal *Cancer Cell*, and journal *Cancer Cell* is assigned to Biochemistry, Genetics and Molecular Biology and Medicine, then we counted "1" for each of the two disciplines.

---
[6] https://www.cortext.net/

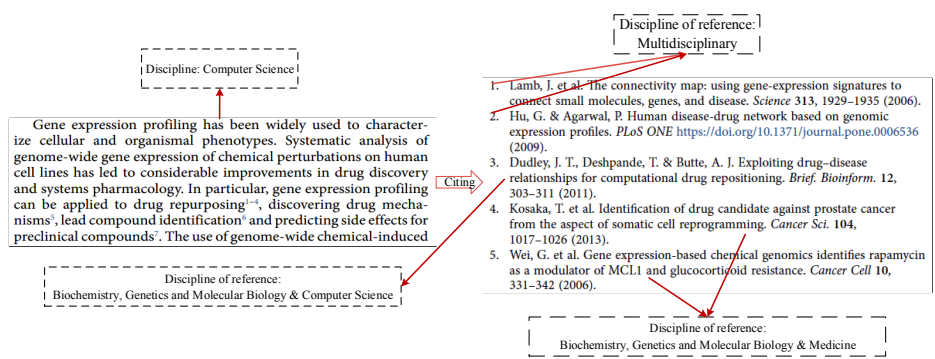

Figure 4 Example of interdisciplinary phenomena in COVID-19 related research

Alan L. Porter and Rafols (2009) pointed out that to characterize the degree of interdisciplinarity, we need to consider multiple aspects, including variety, balance and disparity. Moreover, the $^2D^S$ index is a comprehensive measure that aggregates the aforementioned aspects (L. Zhang et al. 2015) and is widely used to measure the degree of interdisciplinarity (Deng and Xia 2020; Leydesdorff et al. 2016). To provide a comprehensive panorama of interdisciplinarity in the field, all indexes are used in this study.

（1）**Variety index**

Variety indicates that the number of disciplines cited by the coronavirus-related research (Porter. et al. 2007). The formula is described as follows:

$$V = \sum_i SC_i \quad (1)$$

Where $i$ is the $i_{th}$ discipline in the Scopus classification system. $SC_i$ is 1, if the discipline $i$ appears in a reference cited by a coronavirus-related publication, and 0 otherwise. The value of V ranges between 1 and the number of SCs (27 in this study). In this study, we use the average disciplinary variety to perform the analysis. Assuming no differences in discipline status, the greater variety suggests that the coronavirus-related research field is more varied.

（2）**Balance index**

Balance indicates the evenness of the disciplinary distribution. The Gini coefficient is well known for its ability to measure concentration(Lerman and Yitzhaki 1984), and Abramo et al. (2018) pointed out that 1-Gini can be considered a measure of balance. The formula is described as follows:

$$B = 1 - \frac{\sum_i (2i-V-1)x_i}{V \sum_i x_i} \quad (2)$$

Where V denotes the number of disciplines cited by coronavirus-related research, $x_i$ is the number of $i_{th}$ disciplines, i takes values ranging from 1 to V. According to the value of $x_i$, all disciplines are ranked in ascending order. For example, there are three disciplines are cited 10, 22, and 5 times respectively; after sorting, the sequence is 5, 10, 22. The value of V is 3, so the degree of Balance is approximately 0.694. Besides, the higher the degree of balance, the more even the distribution of disciplines.

（3）**Disparity index**

Disparity is defined as the degree to which these disciplines are dissimilar from each other, and disparity is the opposite side of similarity(Alan L. Porter and Rafols 2009). The formulas are described as follows:

$$d_{ij} = 1 - S_{ij} \quad (3)$$

$$S_{ij} = \cos(x_i, x_j) = \frac{\sum_n x_{in} \sum_n x_{jn}}{\sqrt{\sum_n x_{in}^2} \sqrt{\sum_n x_{jn}^2}} \quad (4)$$

Where $S_{ij}$ denotes the degree of similarity between categories i and j. In this study, the cosine similarity method is applied to compute disciplinary similarity according to the disciplinary citation matrix of the references. The degree of disparity falling between 0 and 1. Additionally, greater disparity indicates greater difference between disciplines.

（4）**True diversity**

The $^2D^s$ index is an True Diversity (TD) indicator of interdisciplinarity measurement, and takes variety, balance, disparity into account (L. Zhang et al. 2015), the formula can be described as follow:

$$TD = \frac{1}{\sum (1-d_{ij})(p_i p_j)} \quad (5)$$

Where $p_i$ and $p_j$ represent the proportion of discipline i and j in all the references, $d_{ij}$ is the disparity of discipline i and j. If all references of coronavirus-related article only belong to one discipline, then the value of the TD is 1. In general, the paper's references are associated with more than one discipline. For example, paper A has 10 references, and the references are assigned to discipline a, b, and c. Disciplines a, b, and c are referred to 1, 4 and 5 times respectively, assuming that the degree of disparity between disciplines a and b is 0.4, the degree of disparity between disciplines a and c is 0.5, the degree of disparity between disciplines b and c is 0.6, so the TD of paper A is about 6.211.

## 4 Results

In this section, we first analyze the disciplinary distribution of coronavirus-related research and the disciplinary distribution of the references cited by coronavirus-related research, and then examine the evolutionary trend of interdisciplinary collaboration. Finally, we measure the degree of interdisciplinarity using multiple indicators.

### 4.1 Disciplines involved in coronavirus-related research

We answer RQ1 in this section. Table 2 presents the distribution of disciplines and the result is sorted by the disciplinary frequency of the references in descending order.

As Table 2 shows, regardless of disciplinary distribution of articles(citing) or references,

Medicine (MEDI), Immunology and Microbiology (IMMU), Biochemistry, Genetics and Molecular Biology (BIOC) have the highest percentage among the 27 disciplines, ranking in the top three. Their proportion is as high as 72.22% (2,445,744/3,386,497) from the disciplinary distribution of articles(citing). Likewise, their proportion is over 74% (56,087/75205) from the disciplinary distribution of references. Moreover, only a small proportion of the knowledge used to solve coronavirus-related problems comes from Earth and Planetary Sciences (EARTH) and Energy (ENERGY), and their proportions are less than 0.1% respectively. From the perspective of subject areas, life sciences and health sciences play a major role in tackling the coronavirus related problems, the sum of their disciplinary frequency accounts for 86.90% (65356/75205) of the total articles' (citing) frequency and 85.97% (2911308/3386497) of the total references' frequency, respectively. The distribution of disciplines suggests that coronavirus-related research is heavily dependent on several dominant disciplines, and also illustrates the uneven distribution of disciplines in this research field.

Table 2 Disciplinary distribution of articles(citing) and references (1990-2020)

| No. | Subject Areas | Discipline | Abbreviation | # of articles | # of references |
|---|---|---|---|---|---|
| 1 | Health Sciences | Medicine | MEDI | 32,657 | 1,200,050 |
| 2 | Life Sciences | Immunology and Microbiology | IMMU | 12,535 | 632,028 |
| 3 | Life Sciences | Biochemistry, Genetics and Molecular Biology | BIOC | 10,895 | 613,666 |
| 4 |  | Multidisciplinary | MULT | 2,394 | 185,261 |
| 5 | Life Sciences | Agricultural and Biological Sciences | AGRI | 1,529 | 176,794 |
| 6 | Life Sciences | Pharmacology, Toxicology and Pharmaceutics | PHARM | 3,018 | 111,051 |
| 7 | Health Sciences | Veterinary | VETE | 1,655 | 68,894 |
| 8 | Life Sciences | Neuroscience | NEUR | 1,788 | 67,701 |
| 9 | Physical Sciences | Chemistry | CHEM | 1,499 | 64,424 |
| 10 | Physical Sciences | Chemical Engineering | CHEME | 945 | 33,679 |
| 11 | Physical Sciences | Environmental Science | ENVI | 793 | 27,797 |
| 12 | Physical Sciences | Engineering | ENGR | 579 | 26,504 |
| 13 | Health Sciences | Nursing | NURS | 756 | 26,449 |
| 14 | Physical Sciences | Computer Science | CS | 826 | 21,977 |
| 15 | Physical Sciences | Physics and Astronomy | PHYS | 374 | 21,268 |
| 16 | Physical Sciences | Materials Science | MATER | 405 | 19,698 |
| 17 | Social Sciences | Social Sciences | SOCI | 516 | 18,690 |
| 18 | Physical Sciences | Mathematics | MATH | 309 | 18,119 |
| 19 | Social Sciences | Psychology | PSYC | 606 | 15,091 |
| 20 | Health Sciences | Health Professions | HEAL | 374 | 11,931 |
| 21 | Social Sciences | Business, Management and Accounting | BUSI | 98 | 6,274 |

| 22 | Social Sciences | Arts and Humanities | ARTS | 226 | 5,780 |
|---|---|---|---|---|---|
| 23 | Social Sciences | Economics, Econometrics and Finance | ECON | 114 | 3,824 |
| 24 | Social Sciences | Decision Sciences | DECIS | 39 | 2,913 |
| 25 | Health Sciences | Dentistry | DENT | 149 | 2,744 |
| 26 | Physical Sciences | Energy | ENERGY | 63 | 2,103 |
| 27 | Physical Sciences | Earth and Planetary Sciences | EARTH | 63 | 1,787 |

### 4.2 The evolution of the disciplinary co-occurrence of the references in coronavirus-related research

We employ the Cortext platform to visualize the disciplinary co-occurrence of the references in coronavirus-related research (i.e., to answer RQ2). In order to more clearly display the visualization results, we divide the time span into two segments, 1990-2004 and 2005-2020[7]. As for the parameter settings, we set the number of nodes to be 500, and set the time slots as 15 for each time segment. For the remaining arguments, we use the recommended value of the Cortext platform and the results of consulting with the platform's core contributor (Lionel Villard). The evolution of disciplinary co-occurrence patterns is displayed in Figure 5 and Figure 6. As illustrated in Figure 5 and Figure 6, the bar represents a special interdisciplinary community, the interdisciplinary community is distinguished by the color of the bar, and the bar's height represents the relative size of the interdisciplinary community. The label (e.g., MEDI&IMMU) is the most representative co-occurrence pattern in an interdisciplinary community. Moreover, the continuity of the bar crossing the years means the developmental continuity of the interdisciplinary research.

As Figure 5 shows, from 1990 to 2004, the patterns of co-occurrence between disciplines are relatively diverse. The majority of co-occurrence patterns lack continuity and only exist in a certain year, and the size of the corresponding interdisciplinary community is small. As we can see, these transient co-occurrence relationships are present between distant disciplines, such as Health Professions & Environmental Science (HEAL&ENVI) in 1990, Environmental Science & Mathematics (ENVI&MATH) in 1993, and Engineering & Chemistry (ENGR&CHEM) in 1994. There is a relatively stable evolutionary stream (red bar) across this time period, which mainly covers the disciplines of Immunology and Microbiology (IMMU), Medicine (MEDI) and Biochemistry, Genetics and Molecular Biology (BIOC), and these disciplines are affinitive. At the same time, the size of this interdisciplinary community is growing year by year. Since 1996, a new evolutionary stream (yellow bar) can be observed, and Social Sciences (SOCI) is the dominant discipline in this stream. This relationship stream has diverged and merged many times over the years.

---

[7] If the time span is large, the labels in the Figure will overlap with each other to a large extent, which affects the visualization, so we divide the time span into two time periods, and this decision will not affect the patterns of disciplinary co-occurrence.

In 2000, it split into two interdisciplinary communities, and the two interdisciplinary communities split and fused again in the following years, and finally stabilized. It is noteworthy that there are three interdisciplinary communities in existence as of 2004, including Medicine & Immunology and Microbiology (MEDI& IMMU), Health Professions & Environmental Science (HEAL&ENVI), Engineering & Chemical Engineering (ENGR&CHEME). Interestingly, Health Professions and Environmental Science belong to different subject areas, but these two disciplines are fused into an interdisciplinary community. The possible explanation is that the coronavirus directly or indirectly affects various aspects of life, such as environment, human health etc., which facilitates the collaboration of these two disciplines and the issues can be solved systematically through the fusion of knowledge of those two disciplines.

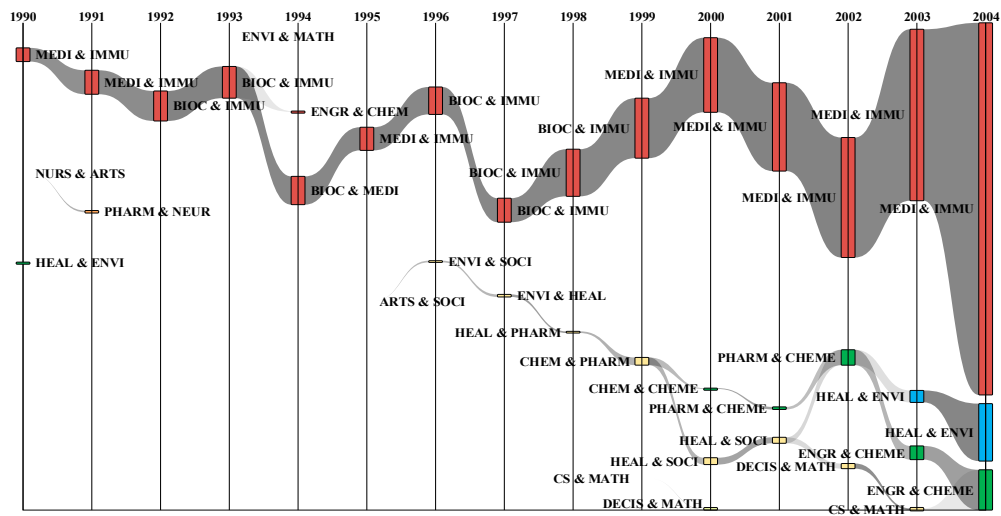

Figure 5 The evolution of disciplinary co-occurrence of the references in coronavirus-related research during 1990-2004. (Note: please refer to Table 2 for the full disciplinary titles of the abbreviations)

From 2005 to 2020, there are three evolutionary streams. Compared to 1990-2004, the three evolutionary streams are very stable, and there has been no frequent splitting and merging of the streams. The red bar represents the Medicine & Immunology and Microbiology (MEDI& IMMU) community, and the size of the community increases year by year. The green bar represents the second type of interdisciplinary community, and the evolutionary stream comprises two major disciplinary co-occurrence relationships, Chemistry & Chemical Engineering (CHEME&CHEM) and Engineering & Chemical Engineering (ENGR&CHEME), most of which are Chemistry & Chemical Engineering (CHEME&CHEM). The blue bar represents the third type of interdisciplinary community, and the stream mainly includes Social Sciences (SOCI), Environmental Science (ENVI), Health Professions (HEAL), Computer Science (CS) and Mathematics (MATH). Social Sciences (SOCI) and Environmental Science (ENVI) act as a bridge between the different disciplinary communities. Overall, the most momentous and longest-lasting disciplinary co-occurrence relationship in the coronavirus-related field is Medicine & Immunology and Microbiology (MEDI& IMMU). Although the co-occurrence relationships between disciplines in coronavirus-related research domain are primarily centered on neighboring disciplines, there are also some relationships between far-flung disciplines such as Computer Science & Environmental Science (CS&ENVI), Mathematics & Environmental Science (MATH&ENVI), which may be related to the attribution of disciplines. For example, Computer Science is the academic discipline concerned with computing, and this discipline can provide tools for other fields. Mathematics is a basic discipline that can provide methods for other fields. Meanwhile, evolutionary streams may be split and merge at different periods, and the change may indicate the exploration of subdivided research fields. Additionally, such frequent splitting and merging of evolutionary stream mainly appeared before 2004.

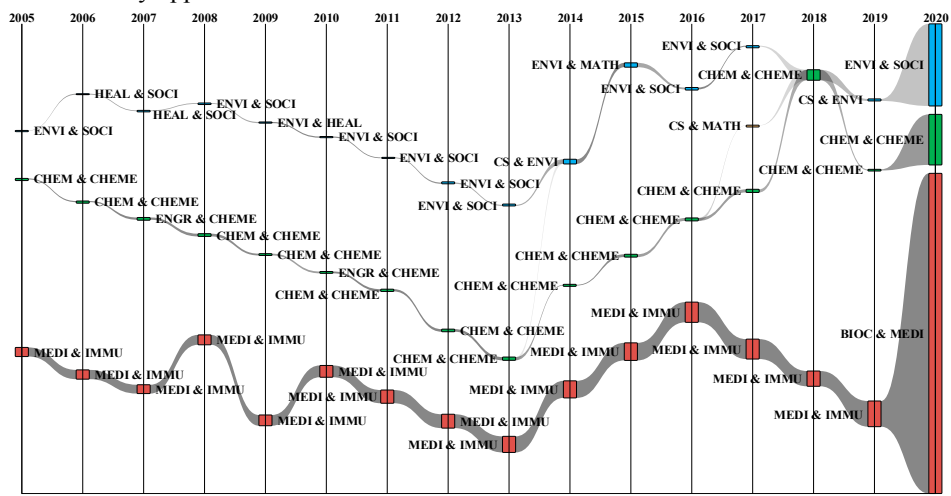

Figure 6 The evolution of disciplinary co-occurrence of the references in coronavirus-related research during 2005-2020. (Note: please refer to Table 2 for the full disciplinary titles of the abbreviations)

### 4.3 Degree of interdisciplinarity in coronavirus-related research

In order to answer RQ3, we use multiple indicators to measure the degree of interdisciplinarity in the coronavirus field. To further explore whether the COVID-19 pandemic accelerates the interdisciplinary degree of this field, we calculate the interdisciplinary degree of coronavirus-related research from 1990 to 2020, and the interdisciplinary degree of COVID-related research in 2020, respectively.

**(1) Disciplinary variety of coronavirus-related research**

As shown in Figure 7, the annual mean number of disciplines shows an intense upward trend before 2020, followed by a dramatic decline in 2020. The annual average number of disciplines increased from 5.471 to 8.067 (about 47% growth) between 1990 to 2019, and the dropped to 6.833 in 2020. From the aspect of disciplinary variety, these findings reveal that coronavirus-related research fields are becoming more diverse before 2020. Scholars tend to absorb knowledge from multiple disciplines to conduct research. Interestingly, there was a sudden drop in the average number of disciplines in 1992, 2002 and 2020. At those certain time points, several coronaviruses-related events occurred: first identification of MERS-Cov in dromedary camels in 1992 (Corman et al. 2014), SARS (Severe Acute Respiratory Syndrome) epidemic during 2002-2003, COVID-19 pandemic in 2020. During an(a) epidemic(pandemic), the interest in coronavirus reemerged as a hot topic. At the same time, to meet the needs of emergency response, scholars need to limit the coordination costs of research (Fry et al. 2020), and utilize their narrow expertise to conduct the research.

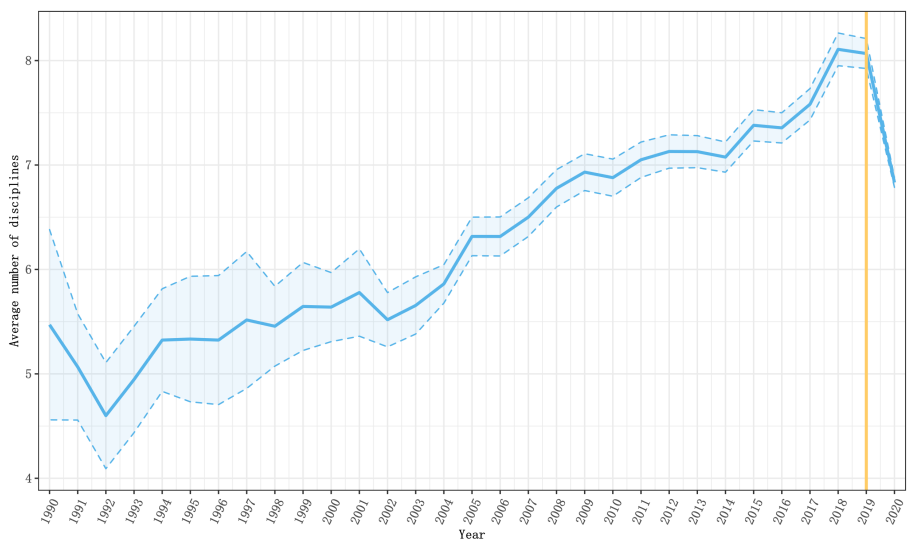

Figure 7 Average number of disciplines in coronavirus-related research from 1990-2020. Shaded areas represent 95% confidence intervals.

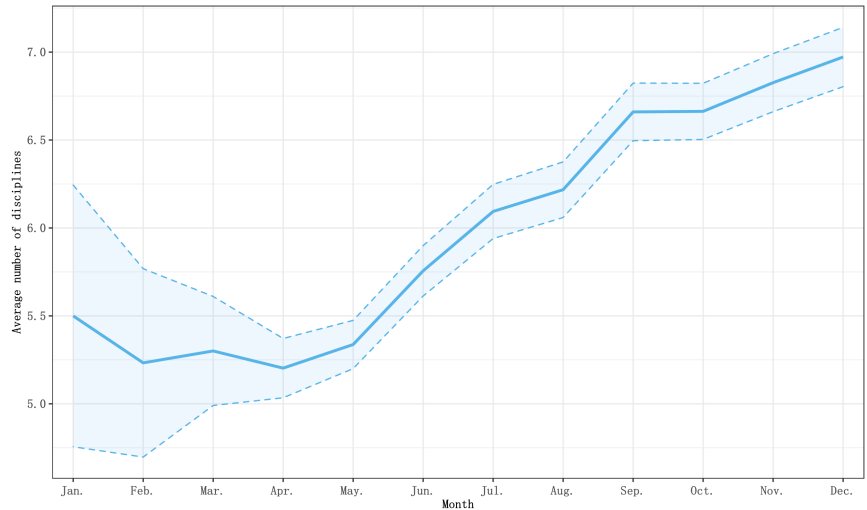

Figure 8 Average number of disciplines of COVID-19 related research in 2020. Shaded areas represent 95% confidence intervals.

The monthly average number of disciplines in 2020 is shown in Figure 8. The monthly average number of disciplines fell from 5.500 in January to 5.203 in April, then rose to 6.972 in December. Moreover, the average number of disciplines of COVID-19 related research is 6.209 in 2020. In comparison with 2019 (8.067), it has reduced by approximately 23%. To a certain extent, this suggests that the COVID-19 pandemic did not accelerate the degree of interdisciplinarity in this field. The possible explanation is that learning about knowledge from other disciplines requires a certain amount of time and cost. Given the suddenness and urgency of the COVID-19 pandemic, scholars need to use their expertise to conduct rapid research, especially those in the public health related fields.

**(2) Disciplinary balance of coronavirus-related research**

As Figure 9 shows, the average disciplinary balance shows a fluctuating decreasing trend. Before 2005, the amplitude of the fluctuations was large, ranging from 0.505 in 1997 to 0.569 in 2002. After 2005, the average disciplinary balance steadily dropped to 0.491 in 2019, and then rose to 0.501 in 2020. The results indicate that the distribution of disciplines is gradually moving toward unevenness. This phenomenon is consistent with the results in the co-occurrence analysis section where the field is dominated by several particular disciplines.

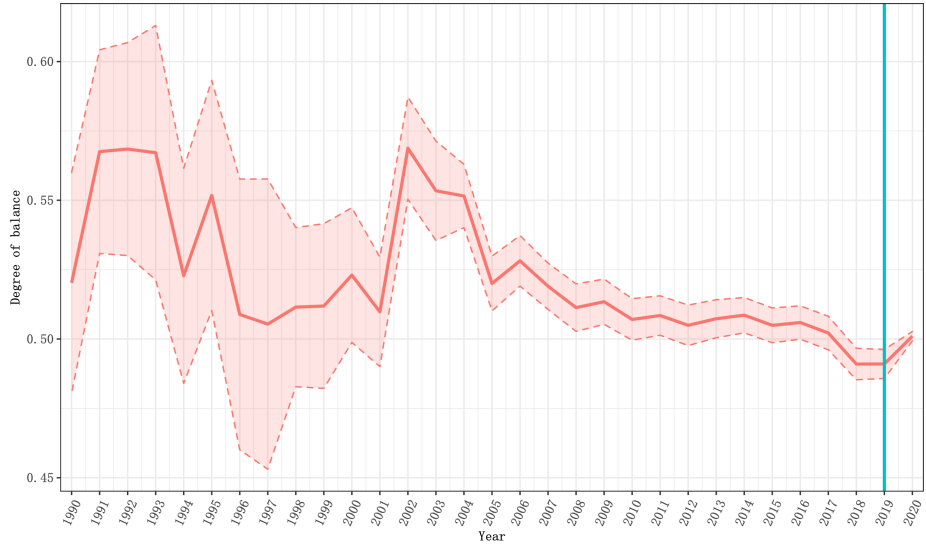

Figure 9 Degree of disciplinary balance of coronavirus related research from 1990 to 2020. Shaded areas represent 95% confidence intervals.

As Figure 10 shows, the degree of disciplinary balance of COVID-19 related research rose from 0.536 in January to 0.587 in February, and then fell to 0.488 in December. In addition, the average degree of disciplinary balance is 0.515 in 2020, which is higher than it was in 2019. The possible explanation is that research focus primarily on public health issues at the beginning of the COVID-19 pandemic, and only a few disciplines are engaged in this field, such as Medicine, Immunology and Microbiology. Based on this logic, we can hypothesize that as the pandemic progresses, more disciplines will be involved in this field (Bontempi et al. 2020), the number of disciplines will increase gradually, and the degree of disciplinary balance will decrease.

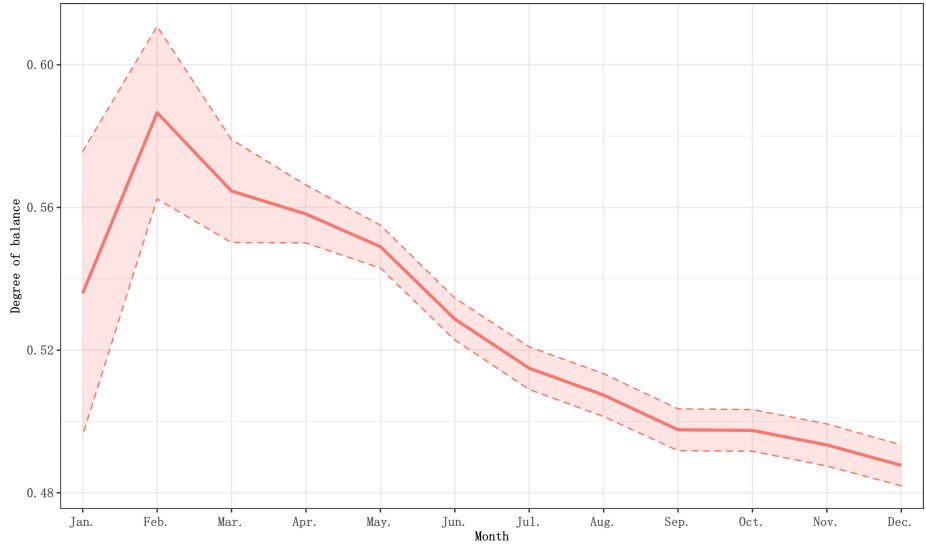

Figure 10 Degree of disciplinary balance of COVID-19 related research in 2020.
Shaded areas represent 95% confidence intervals.

**(3) Disparity between disciplines in coronavirus-related research**

Using formulas (3) and (4), we calculate the similarity $S_{ij}$ between disciplines, and then obtain the degree of difference between disciplines (Deng and Xia 2020; X. Xu et al. 2021). As can be seen in Figure 11, there is a significant disparity between Decision Sciences (DECIS) and Veterinary (VETE). Additionally, the disciplines of Business, Management and Accounting (BUSI) and Decision Sciences (DECIS) are quite different from the other disciplines. We also find that Medicine (MEDI), Agricultural and Biological Sciences (AGRI), Biochemistry, Genetics and Molecular Biology (BIOC), Immunology and Microbiology (IMMU) have relatively high levels of disciplinary similarity, which are above 90%, indicating that the research fields of these disciplines are relatively close.

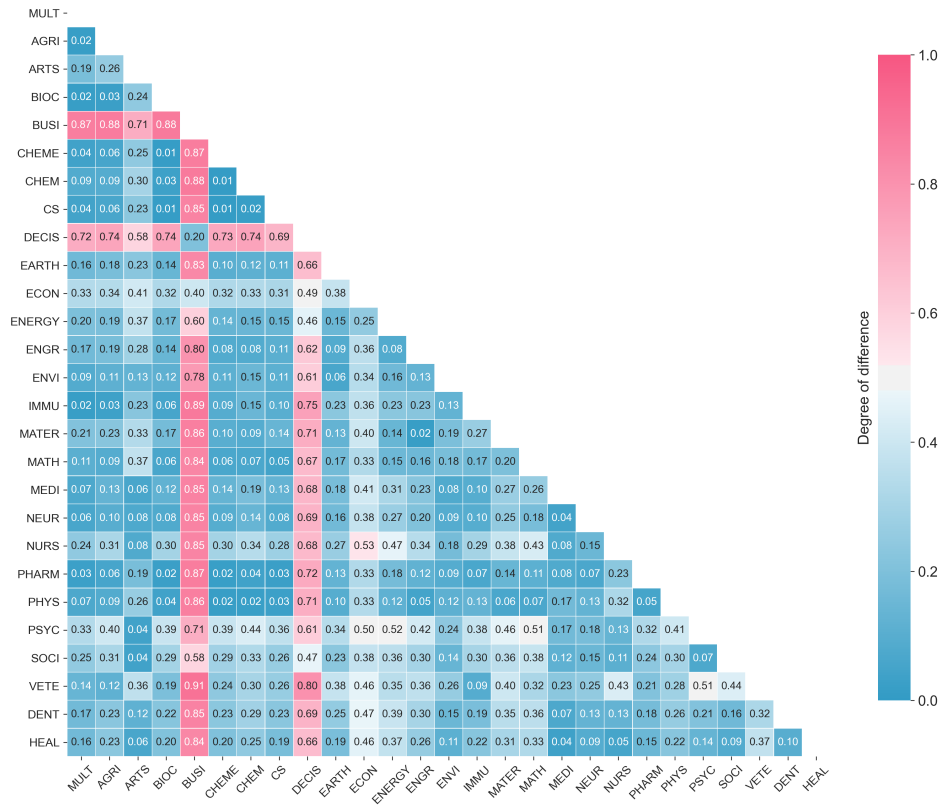

Figure 11 Disparity between disciplines in coronavirus-related research. (Note: please refer to Table 2 for the full disciplinary titles of the abbreviations)

**(4) The degree of interdisciplinarity in coronavirus-related research**

The $^2D^s$ index is also a comprehensive indicator for quantifying interdisciplinary research. As shown in figures 12 and 13, we calculate the average integrated diversity for the two time periods 1990-2020 and January-December 2020. We can see in Figure 12 that the average integrated diversity fluctuates in the early years, after 2002, it shows a slow downward trend, and then it rises sharply in 2020, and the average integrated diversity increased by approximately 39% compared to 2019. Moreover, there are two peaks in Figure 12. The first peak is in 1992, while the second is in 2001, and the average integrated diversity are 7.350 and 6.061, respectively.

As Figure 13 shows, the average degree of true diversity shows a fluctuating declining trend in 2020. The average degree of true diversity reaches a nadir (4.757) in February. Moreover, average integrated diversity of COVID-19 related research is 5.378, which is higher than the average integrated diversity of coronavirus-related research in 2019. The above results suggest that the coronavirus-related research is becoming less interdisciplinary before COVID-19 pandemic, and the COVID-19 pandemic makes the coronavirus research field more interdisciplinary. In section 3.2, we can visually observe that the disciplinary co-occurrence relationship is mostly present in neighboring disciplines, such as Immunology and Microbiology, Medicine, Biochemistry, Genetics and Molecular Biology, which reduces the disparity between disciplines and increases the corresponding denominator of the formula (5), resulting in the values of true diversity decreasing. The results also support the conclusions in section 4.2, by observing the diachronic evolution of the disciplinary co-occurrence relationship, we find that several disciplines in this field merge into three stable communities and dominate the coronavirus-research field after 2004, we can infer that degree of interdisciplinarity is decrease over time based on this observed result. As for the reason why COVID-19 accelerated the degree of interdisciplinarity, one possible explanation is that high priority research such as virus structure analysis and patient's clinical manifestations are conducted in the early stages of the COVID-19 pandemic, and these topics do not require a wide range of interdisciplinary knowledge. Therefore, this may result in a highly uneven distribution of disciplines, and decreases the denominator of the formula (5), leading to the degree of true diversity increasing.

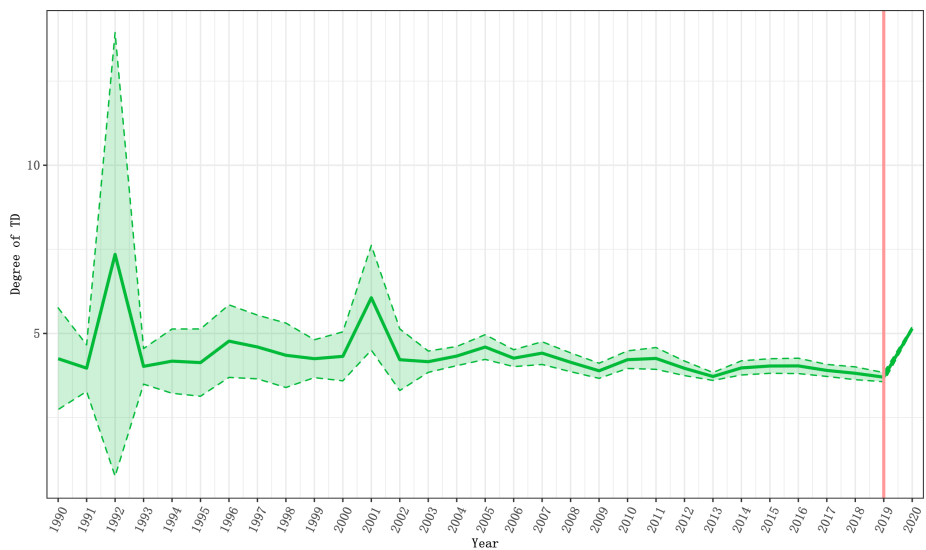

Figure 12 The True Diversity in coronavirus related research from 1990 to 2020. Shaded areas represent 95% confidence intervals.

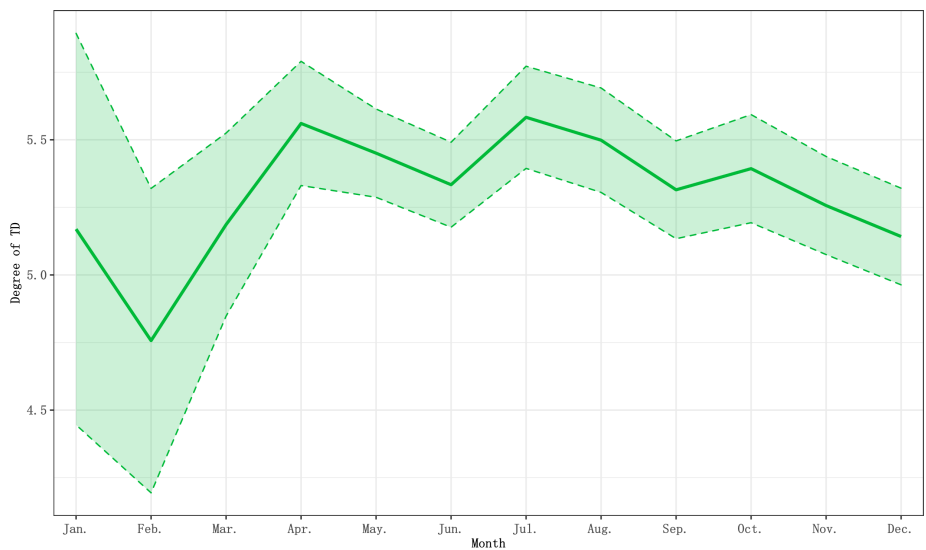

Figure 13 The True Diversity of COVID-19 related research in 2020. Shaded areas represent 95% confidence intervals.

## 5   Discussion

This study investigates the disciplinary distribution of the coronavirus-related research and the disciplinary distribution of references cited by the coronavirus-related research, and explores the evolutionary trends of the disciplinary co-occurrence relationship and calculates the interdisciplinary degree of this field in the past 31 years. Furthermore, we demonstrate that the coronavirus-related research is not becoming more interdisciplinary before 2019, and the COVID-19 pandemic makes the coronavirus research field more interdisciplinary. These results are helpful to obtain an interdisciplinary picture as complete as possible in the field of coronavirus related research.

### 5.1   Consolidating the status of the backbone discipline of coronavirus research

The coronavirus-related research covers an increasing number of disciplines before 2019, but this field centers on a few dominant disciplines, including Medicine, Immunology and Microbiology, Biochemistry, Genetics and Molecular Biology. Through a co-occurrence analysis, we discover that these three disciplines are frequently co-occurring. Interdisciplinary knowledge transfer is more easily occurring between neighboring disciplines (Morillo et al. 2003). This may be leading to the decrease of interdisciplinary degree. Owing to the disciplinary identity, different disciplines have different intellectual culture and research priority. The primary mission of immunology and microbiology is to study the transformation and outcome of coronavirus, while clinical medicine is mainly to observe the changes in patients' condition and develop therapeutic strategies. Basic research in Immunology and Microbiology can offer theoretical support for clinical research in Medicine. Disciplines have been institutionalized through processes of specialization, and specialization is the prerequisite to build a discipline, but it should not become a religion (Serpa et al. 2017). At the same time, interdisciplinarity is hard to be implemented, so it is vital to protect the relationships between disciplines that have been connected. Given the high efficiency of communication between neighboring disciplines, policymakers need to consciously establish several interdisciplinary laboratories or funds for these closely disciplines to consolidate the backbone status of this field, thereby bringing continuous contributions to the development of coronavirus related research.

### 5.2 Interdisciplinary research provides systematic solutions for the COVID-19 pandemic

The coronavirus-related research is not becoming more interdisciplinary before 2019. In contrast to 2019, the average number of disciplines of COVID-19 related research decreased in 2020, while the degree of interdisciplinarity increased in 2020. Science rapidly shifted to engage COVID-19 following the emergence of virus, and understanding the research questions related to the biology of this virus must be a top priority, which mainly take advantage of the knowledge from biomedical-related disciplines. As a result, the distribution of disciplines is more balanced and concentrate on certain disciplines, resulting and increase in true diversity. COVID-19 pandemic has dramatically impacted all spheres of human life, and many scientists have called on scholars to conduct interdisciplinary research (Holmes et al. 2020; Bontempi et al. 2020). Interdisciplinary research is related to creativity and innovation and a lot of intellectual "breakthroughs" are acquired by crossing disciplinary boundaries (Morillo et al. 2003). Although many scientists across all fields have jumped into the COVID-19 research streams (Hill et al. 2021), the monthly true diversity is decreasing, which may indicate that the real interdisciplinary solution is lacking. Using fake news detection during COVID-19 as an example, computer scientists can easily apply machine learning techniques to identify fake news (Khanday et al. 2020), but it is not the key issue that misinformation brings to society. The acceptance of the views expressed by fake news is associated with various factors, including level of education, the trust in expertise, etc. The detection of fake news cannot completely prevent the diffusion of fake news. It is therefore essential to integrate the knowledge of Computer Science, Social Science, Psychology and other disciplines to generate a systematic solution. However, conducting interdisciplinary research is not that easy; Compared to what happens in the consolidated scientific disciplines, the results of interdisciplinary studies are more uncertain. It is crucial for policymakers and research managers to provide institutional and administrative support for scholars to foster interdisciplinary research.

### 5.3 Accelerating the collaboration among distant disciplines in coronavirus-related research

The results of the co-occurrence analysis reveal that the neighboring disciplines are more likely to appear in the same publication, but the innovation of this field can also be fueled by the knowledge from the distant disciplines (Yoo 2017). Some patterns of disciplinary co-occurrence are just a flash in the pan, including Health Professions & Environmental Science, Mathematics & Environmental Science, Computer Science & Environmental Science. The connections between these distant disciplines are tenuous and easy to diverge or merge. The possible reason is that the cognitive barriers between distant disciplines are relatively high. Although the brief appearance of co-occurrence relationships among these distant disciplines is natural, it does not favor the enrichment of the knowledge system of the coronavirus-related field, and hinders the generation of new knowledge or theories. Sometimes, breakthrough discoveries are missed just because researchers fail to make the link between disciplines (Chai 2017), so efforts are required to reduce the communication threshold between distant disciplines. One solution is to build the scholars' profile for all scholars who conduct the coronavirus-related research, and integrate the collaboration network, relationships between scholars and knowledge entities, and scholars' personal information into it (J. Xu et al. 2020). Another solution is to establish a regular forum for academic communication between distant disciplines (Cummins et al. 2018). Taking the disciplinary collaboration between medicine and engineering as an example, the interdisciplinary forum provides a means of formal knowledge exchange in which engineers are informed of medical challenges and relevant technologies are highlighted to doctors.

## 6 Conclusion and Future Works

Based on several indicators and co-occurrence analysis methods, this study has presented an intuitive understanding of the interdisciplinarity of coronavirus-related research between 1990 and 2020. Our analysis demonstrates that biomedical-related disciplines are the dominant disciplines in coronavirus-related research, including Medicine, Immunology and Microbiology, Biochemistry, Genetics and Molecular Biology. In terms of the co-occurrence relationships, Immunology and Microbiology & Biochemistry, Genetics and Molecular Biology (IMMU&BIOC) and Multidisciplinary & Biochemistry, Genetics and Molecular Biology (MULT&BIOC) are two stable types of co-occurrence relationships and have continued throughout the whole research period. Additionally, there are several transient relationships, such as Engineering & Chemistry (ENGR&CHEM)，Environmental Science & Mathematics (ENVI&MATH). When it comes to degree of interdisciplinarity, the number of disciplines increases, while the disciplinary balance and true diversity indicators show a decreasing trend from 1990 to 2019. By analyzing the COVID-19 related papers published between January and December 2020, the disciplinary variety reveal an overall upward trend by month, while disciplinary distribution and true diversity show a fluctuating decline trend by month. Furthermore, the average true diversity of coronavirus-related research in 2020 is higher than it was in 2019. The above findings suggest that the coronavirus-related research is not becoming more interdisciplinary, but the COVID-19 pandemic accelerate the degree of interdisciplinarity.

There are some limitations in this work. First, a few articles lack corresponding references due to technical issues, which inevitably lead to a bias in these data. Consequently, researchers can consider integrating multiple data sources such as Scopus, Web of Science, PubMed, etc., and developing an article metadata extraction framework to extract the relevant metadata, and then a more accurate and complete understanding of interdisciplinarity in coronavirus related research can be conducted. Second, the disciplinary classification standards used in this study are drawn from the Scopus discipline classification system, and disciplines are determined by the journal discipline, but the disciplinary classification standards of different bibliographic databases are not universal (Hu and Zhang 2017; Alan L. Porter and Rafols 2009), and the use of different classification standards may lead to different results for understanding the interdisciplinarity of the field. In addition, there are limitations to determining the discipline of papers based on the discipline of journals. Although this method is widely adopted in bibliographic analysis (C. Zhang et al. 2021; Deng and Xia 2020), the discipline of journals sometimes does not accurately represent the discipline of the papers. For example, a paper may be assigned to an unrelated discipline simply because the journal to which it belongs has this discipline. This will introduce bias into the calculation of interdisciplinarity indicators. Furthermore, only articles indexed in Scopus can be assigned to one or more disciplines. Specifically, other types of articles are not allocated to any disciplines. This likely reduces the number of articles available, and may bring some bias into the results.

There are many related studies that can be implemented in the future. First, this article mainly focuses on a broad view on the interdisciplinarity of coronavirus-related field; a more nuanced investigation should be conducted on interdisciplinary differences between different types of coronavirus research (e.g., SARS-CoV, MERS-CoV, SARS-CoV-2). Second, this work examines the interdisciplinarity of coronavirus research at the level of discipline categories. Citation relationships between publications can be employed to cluster the research fields, and then the degree of interdisciplinarity can be conducted at the level of publication rather than at the level of journal, which will provide a more detailed observation on the flow of knowledge and the interaction of disciplines (Waltman and Eck 2012).

## Acknowledgment

This work is supported by Jiangsu Social Science Fund (Grant No. 20TQA001), and Postgraduate Research & Practice Innovation Program of Jiangsu Province (grant no. KYCX20_0347). The authors would like to thank the anonymous reviewers for their precious comments and suggestions. The authors would also like to thank Chen Yang, Yuntian Song, Liang Tian, from School of Economics and Management, Nanjing

## Author Biography

**Yi Zhao** is currently a PhD student in the School of Economics and Management, Nanjing University of Science and Technology. He received his master's Degree of Economics from Hohai University, Nanjing, China, in 2019. His research interests mainly focus on knowledge organization and text mining. He has published four publications including DIM, DI, ASIS&T, ISSI.

**Lifan Liu** received her bachelor's degree in information management and system program from Nanjing University of Information Science and Technology, Nanjing, China, in 2018. She received her master's degree in information science program from Nanjing University of Science and Technology, Nanjing, China, In 2021. Her research interests include text mining and informetrics. She has published three publications including JOI, ASIS&T.

**Chengzhi Zhang** is a professor of Department of Information Management, Nanjing University of Science and Technology, China. He received his PhD degree of Information Science from Nanjing University, China. He has published more than 100 publications, including JASIST, Aslib JIM, JOI, OIR, SCIM, ACL, NAACL, etc. His current research interests include scientific text mining, knowledge entity extraction and evaluation, social media mining. He serves as Editorial Board Member and Managing Guest Editor for 10 international journals (Patterns, OIR, Aslib JIM, SCIM, TEL, IDD, NLE, JDIS, DIM, DI, etc.) and PC members of several international conferences in fields of natural language process and scientometrics. (https://chengzhizhang.github.io/)